\documentclass[runningheads]{llncs}
\usepackage[T1]{fontenc}
\usepackage{graphicx}
\usepackage{amsmath}
\usepackage{subcaption}

\usepackage{amsfonts}
\usepackage{amssymb}
\usepackage{xcolor}
\usepackage{color}
\usepackage{amsmath}
\usepackage{amsfonts}
\usepackage{amssymb}
\usepackage{esint}
\newcommand{\ie}{\textit{i.e.}, }

\begin{document}
\begin{sloppypar}
%
\title{Suppressing seizure via optimal electrical stimulation to the hub of epileptic brain network}
\titlerunning{Suppressing seizure via optimal electrical stimulation}
%
\author{Zhichao Liang\inst{1}$^{,\dagger}$ \and
Guanyi Zhao\inst{1}$^{,\dagger}$ \and
Yinuo Zhang\inst{1} \and
Weiting Sun\inst{1} \and
Jingzhe Lin\inst{1} \and
Jialin Wang\inst{2} \and
Quanying Liu\inst{1}}
\authorrunning{Z. Liang, G. Zhao et al. 2024}
%
\institute{$^1$ Department of Biomedical Engineering, Southern University of Science and Technology, Shenzhen, CHINA \\
$^2$ Shenzhen Middle School, Shenzhen, China  \\
$^{\dagger}$ Equal contribution \\
Corresponding to \email{liuqy@sustech.edu.cn} (Q.Liu)}
\maketitle              
\begin{abstract}
The electrical stimulation to the seizure onset zone (SOZ) serves as an efficient approach to seizure suppression. Recently, seizure dynamics have gained widespread attendance in its network propagation mechanisms. Compared with the direct stimulation to SOZ, other brain network-level approaches that can effectively suppress epileptic seizures remain under-explored. In this study, we introduce a platform equipped with a system identification module and a control strategy module, to validate the effectiveness of the hub of the epileptic brain network in suppressing seizure. The identified surrogate dynamics show high predictive performance in reconstructing neural dynamics which enables the model predictive framework to achieve accurate neural stimulation. The electrical stimulation on the hub of the epileptic brain network shows remarkable performance as the direct stimulation of SOZ in suppressing seizure dynamics. Underpinned by network control theory, our platform offers a general tool for the validation of neural stimulation.

\end{abstract}

\section{Introduction}
Episodes of focal epilepsy engage networks that operate at various spatiotemporal scales~\cite{laufs2012functional,bartolomei2013abnormal,zaveri2008large}. 
The epileptogenic network (EN), formed by the brain regions that are actively involved in the initiation of epileptic seizures, has gained widespread attention~\cite{bartolomei2017defining,wang2023delineating}. The abnormal activities within specific nodes of epileptogenic networks typically show cascade reactions. Evidence has been reported that EN directly impacts the success of epilepsy surgical interventions~\cite{burns2014network}. Therefore, it is hypothesized that appropriate electrical stimulation to EN may help suppress seizures. 

Nearly one-third of epilepsy patients are drug-resistant, which means medication cannot effectively control their seizures~\cite{rocha2023pharmacoresistance}. Although epilepsy surgery is considered as primary treatment for drug-resistant epilepsy, it has significant risks, such as irreversible structural damage to the brain. Alternatively, electrical stimulation, as an advanced technique for neurodiagnostic and therapeutic purposes, has the merits of safety and flexibility. Some pioneer studies have reported that neural stimulation effectively suppressed epileptic seizures by desynchronizing the epileptogenic networks~\cite{de2014focus,yu2018high,scherer2020desynchronization}. In clinical practice, stimulation targets are typically selected within the EN or some other connected brain areas~\cite{sisterson2020neuromodulation}. However, the regulatory strategies and effects of these regions lack theoretical and experimental quantitative validation~\cite{schaper2023mapping}. More importantly, these stimulation strategies are highly dependent on the clinician's experience rather than systematic quantitative design~\cite{piper2022towards}. There is an urgent need to model optimal stimulation strategies to provide personalized treatment plans for patients~\cite{xia2024controlling}.

The human brain is a complex network-coupled dynamical system that the information transits within the network~\cite{luo2022mapping}. Due to the physical constraints of directly stimulating the target area, we aim to develop a feasible and optimal control strategy at the network level. In the past few years, the network control theory has provided a theoretical framework for the understanding of the controllability of specific nodes~\cite{gu2015controllability,gu2017optimal}. Therefore, from a network perspective, especially in EN, it is worth studying how to select control nodes and whether the optimal control strategy applied to the selected control nodes can effectively suppress the capture dynamics. 

In this study, we aim to design effective control strategies based on network control theory to suppress epileptic seizures in neural activity. To this end, we use a linear dynamical model to reconstruct the neural dynamics of seizure propagation, enabling it to serve as a virtual platform for accurately predicting epileptic activities. Based on this surrogate model, combined with Model Predictive Control (MPC) theory, we calculate the optimal control strategies for different control nodes. 
The main contributions of this study are summarized as follows.
\begin{itemize}
    \item We present an optimal control framework for seizure suppression, which contains a data-driven system identification module and a control strategy design module (Fig.~\ref{fig:framework}).
    \item We demonstrate that the control strategy on the hub of the epileptic brain network, as well as on the SOZ, can effectively suppress seizures with evidence from simulation (Fig.~\ref{fig:JR_Simulation}) and from the real iEEG data (Fig.~\ref{fig:real_data}).
\end{itemize}

\section{Related work}

Some electrical stimulation techniques have been used for seizure suppression, including deep brain stimulation (DBS)~\cite{halpern2009deep}, responsive neurostimulation (RNS)~\cite{morrell2011responsive}, vagus nerve stimulation (VNS)~\cite{morrell2011responsive} and some non-invasive transcranial stimulations~\cite{simula2022transcranial,wang2023multi}. Past research has employed empirical stimulation parameters for long-term stimulation in epilepsy patients~\cite{roa2023long,khambhati2021long}. A major challenge to applying electrical stimulation for seizure suppression lies in choosing the stimulation sites and designing the optimal stimulation parameters. Recent studies have proposed the Koopman-MPC framework~\cite{liang2022online} and unscented Kalman filter~\cite{chang2021data} to optimize the stimulation parameters, such as frequency, intensity, and timing of stimulation. However, theoretical and computational work on choosing stimulation sites is still missing.

Clinically, common targets for epilepsy electrical stimulation include the cerebellum, centromedian thalamus, hippocampus, anterior nucleus of the thalamus, motor cortex, caudate, subthalamic nucleus, and the epileptogenic foci~\cite{fisher2014electrical}. Currently, the selection of stimulation targets is predominantly based on the typical regions of epileptic lesions, a method that lacks objective quantitative metrics and heavily relies on the physician's experiential judgment, making it challenging to tailor personalized treatment strategies for different patients. Therefore, there is a critical need for the development of objective quantitative indices to assess different stimulation nodes quantitatively, thereby optimizing stimulation strategies further.




The consumption of energy during state transition with brain network control theory characterizes the feasibility or difficulty of reaching the target state. Especially in the feasibility of state transition of cognitive function~\cite{gu2015controllability,gu2017optimal,liang2024reverse}. 
Gu et al. proposed a brain network control strategy, modeling the brain using a linear dynamical model to calculate the energy cost of transitions between different brain states from a network perspective~\cite{gu2017optimal}. By quantitatively calculating the optimal energy input required for specific nodes, the nodes that have stronger driving forces on brain states can be identified ~\cite{liang2024reverse}.  Some other studies have applied network control theory to understand the worm movement~\cite{yan2017network}, as well as to guide electrical stimulation~\cite{stiso2019white}.




\section{Method}

\begin{figure}[th]
\centering
\includegraphics[width=\textwidth]{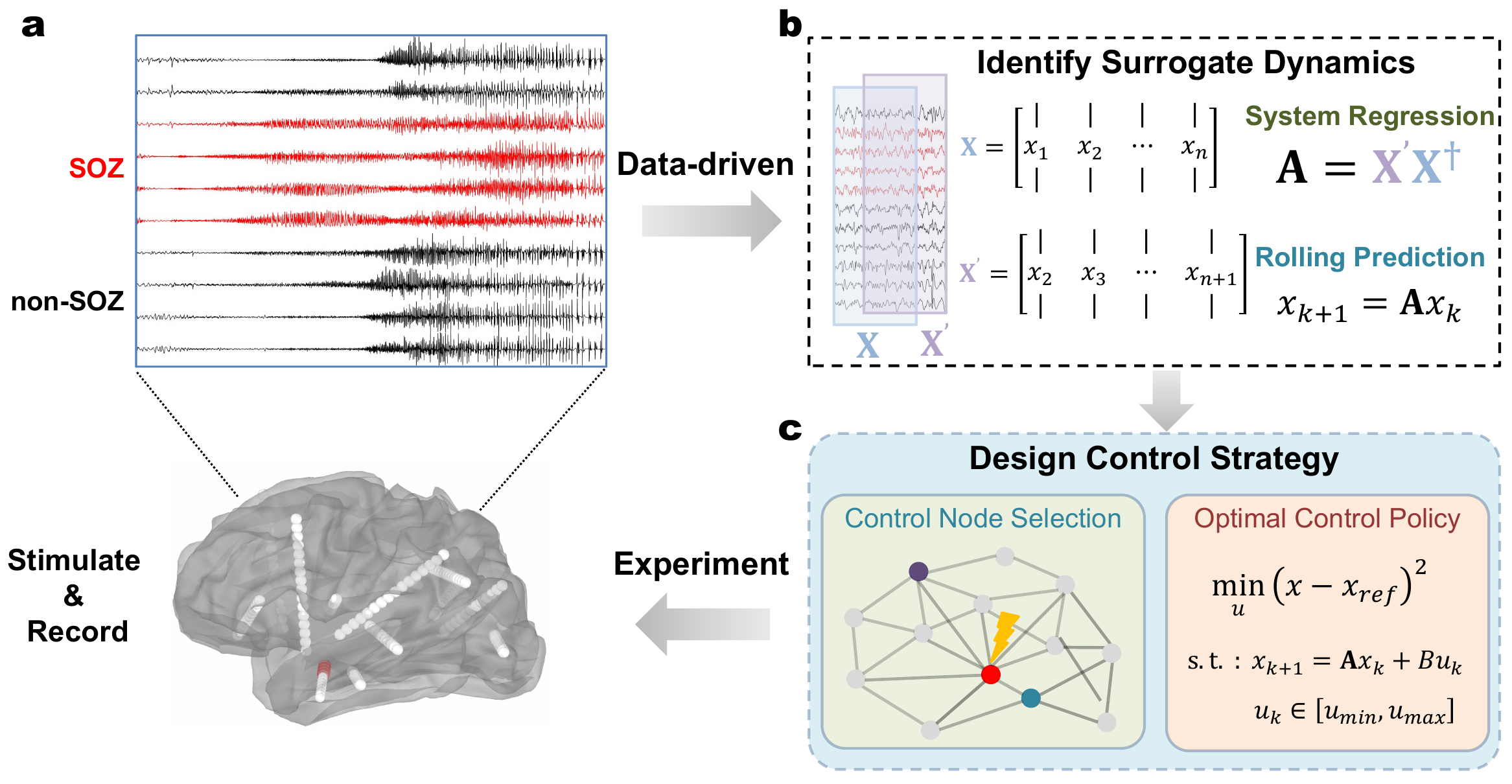}
\caption{\textbf{The optimal control framework for seizure suppression}, including a system identification modular and an optimal control strategy modular. \textbf{a}, Brain nodes in SOZ (red lines) and non-SOZ (black lines) of a subject are pre-determined by clinicians and the real iEEG data is from the Epilepsy-iEEG-Multicenter-Dataset. \textbf{b}, Identifying the system dynamics via reconstructing the neural dynamics. \textbf{c}, Designning the optimal control strategy, considering two factors: control node selection (left) and the optimal control policy (right). The dark nodes indicate potential control nodes. The optimal inputs $u$ to the control nodes are designed by minimizing the error between the surrogate dynamics and the reference dynamics and control energy. } 
\label{fig:framework}
\end{figure}

In this section, we introduce our framework for suppressing seizures via electrical stimulation to the hub of the epileptic brain network (see Fig.~\ref{fig:framework}). The framework is composed of two parts: (1) Learning the surrogate neural dynamics of seizure-like activity (\textbf{System Identification}). (2) Selecting the control node and calculating the optimal stimulation policy to suppress seizure (\textbf{Control Strategy}).

\subsection{Problem definition}
To address the control problem in suppressing seizure dynamics, we need to design an accurate prediction model and a valid control strategy. We treat the neural dynamics of seizure propagation as a dynamical system, which can be approximated with a linear dynamical system or more advanced surrogate models in a data-driven fashion (Fig.~\ref{fig:framework}b). The reconstructed dynamics require the capability of fast online learning and accurate prediction. The control strategy is designed under the guidelines of the surrogate dynamics with the consideration of control node selection and the corresponding optimal stimulation (Fig.~\ref{fig:framework}c). Meanwhile, we need to consider the safety of electrical stimulation. We introduce the range of minimum or maximum stimulation parameters in the optimization problem of obtaining the optimal control strategy. 

\subsection{Dynamic mode decomposition in identifying seizure dynamics}
Dynamic mode decomposition (DMD) 
has been widely applied in extracting the spatio-temporal dynamic modes of resting state and task-related fMRI BOLD signals~\cite{casorso2019dynamic} or predicting behaviour~\cite{ikeda2022predicting}. In our study, to investigate the local neural dynamics from data, we applied DMD to reconstruct seizure dynamics, as shown in Fig.~\ref{fig:framework}b.

From the observed seizure data, we can obtain a sequence of vector data $\{{x}_{1},{x}_{2},\ldots,{x}_{n},{x}_{n+1}\}$, where ${x}_{i}\in \mathcal{R}^{m}$ representing the activity of $m$ recording electrodes at time step $i$. We can construct the following two matrices $X$ and $X^{\prime}$ with: 
\begin{equation}
\begin{aligned}
 X=\begin{bmatrix}|&|&&|\\x_1&x_2&\cdots&{x}_{n}\\|&|&&|\end{bmatrix} \quad \textit{and} \quad 
 X'=\begin{bmatrix}|&|&&|\\ x_2&x_3&\cdots&x_{n+1}\\|&|&&|\end{bmatrix}
\end{aligned}
\end{equation}
where ${X}$ and $X^{\prime} \in \mathcal{R}^{m \times n}$, $n$ is the length of time window (Fig.~\ref{fig:framework}b). 

We can identify the seizure dynamics by using the DMD algorithm, which aims to identify the linear operator $A$ by approximating the linear dynamics of
\begin{equation}
\begin{aligned}
{x}_{t+1}=A{x}_t,
\end{aligned}
\end{equation}
written in terms of these data matrices as
\begin{equation}
\begin{aligned}
{X^{\prime}\approx AX}.
\end{aligned}
\end{equation}
The best fit of linear operator $A$ can be solved by minimizing the Frobenius norm of $\|{X}^{\prime}-{A}{X}\|_F$:
\begin{equation}
\begin{aligned}
{A}={X}^{\prime}{X}^{\dagger}
\end{aligned}
\end{equation}
where ${X}^{\dagger}$ is the pseudo-inverse of $X$.

In DMD theory, to address potential high-dimensional problems and perform dimension reduction, ${X}$ is subjected to Singular Value Decomposition with $X = U\Sigma V^*$, where $*$ denotes complex-conjugate transpose. Then the pseudo-inverse of ${X}$ is ${X}^{\dagger} = {V}{\Sigma}^{-1}{U}^{*}$, and $X^{\prime}=AU\Sigma V^*$, where ${U}\in{R}^{m\times m}$ and ${V}\in{R}^{n\times n}$ are unitary.


\subsection{Control policy}
Model predictive control is an optimization-based control framework. It minimizes the objective function in a finite prediction horizon with the control inputs and states transition constraints. In this study, we design the control strategy by considering the control node selection and its optimal control policy under the model predictive control framework. 
The standard objective function of MPC is defined as follows,
\begin{equation}
\label{Eq:MPC}
    \begin{array}{cl}
    \min\limits_{u} & \sum \limits_{t=1}^{T_p} ( x_{t}- x_{ref})^{T} Q_x( x_{t}- x_{ref}) + \sum \limits_{t=1}^{T_c}u_{t}^T Q_u u_{t}\\
    \text{ s.t. }  & x_{t+1} = f(x_{t},u_{t}) \\
    & u_{t} \in [u_{\min}, u_{\max}]  \\
\end{array}
\end{equation}
where $x_t$ represents the system state and $x_{ref}$ is the reference signal, $u_t$ is the input of the system, $Q_x$ is the positive definite weighted matrix for penalizing the deviance and $Q_u$ is a non-negative matrix for regularizing the amplitudes of control inputs. $T_p$ and $T_c$ are the length of the predictive horizon and control horizon, respectively. 

In control engineering, the method to solve the optimization problem in Eq.~\eqref{Eq:MPC} is chosen based on the system dynamics $f(\cdot)$. If $f(\cdot)$ is a nonlinear function of $x_t$ and $u_t$, it becomes a nonlinear model predictive control problem and can be solved by a nonlinear programming solver with the primal-dual interior method. If $f(\cdot)$ is a linear function of $x_t$ and $u_t$ (\ie $x_{t+1} = Ax_t+Bu_t$), the optimization problem can be solved by a linear quadratic programming solver directly. In both cases, the optimization problem in Eq.~\eqref{Eq:MPC} can also be solved by other gradient descent methods (\ie Adam, SGD).


\subsection{Simulation platform for seizure propagation}

\begin{figure}[ht]
\centering
\includegraphics[width=\textwidth]{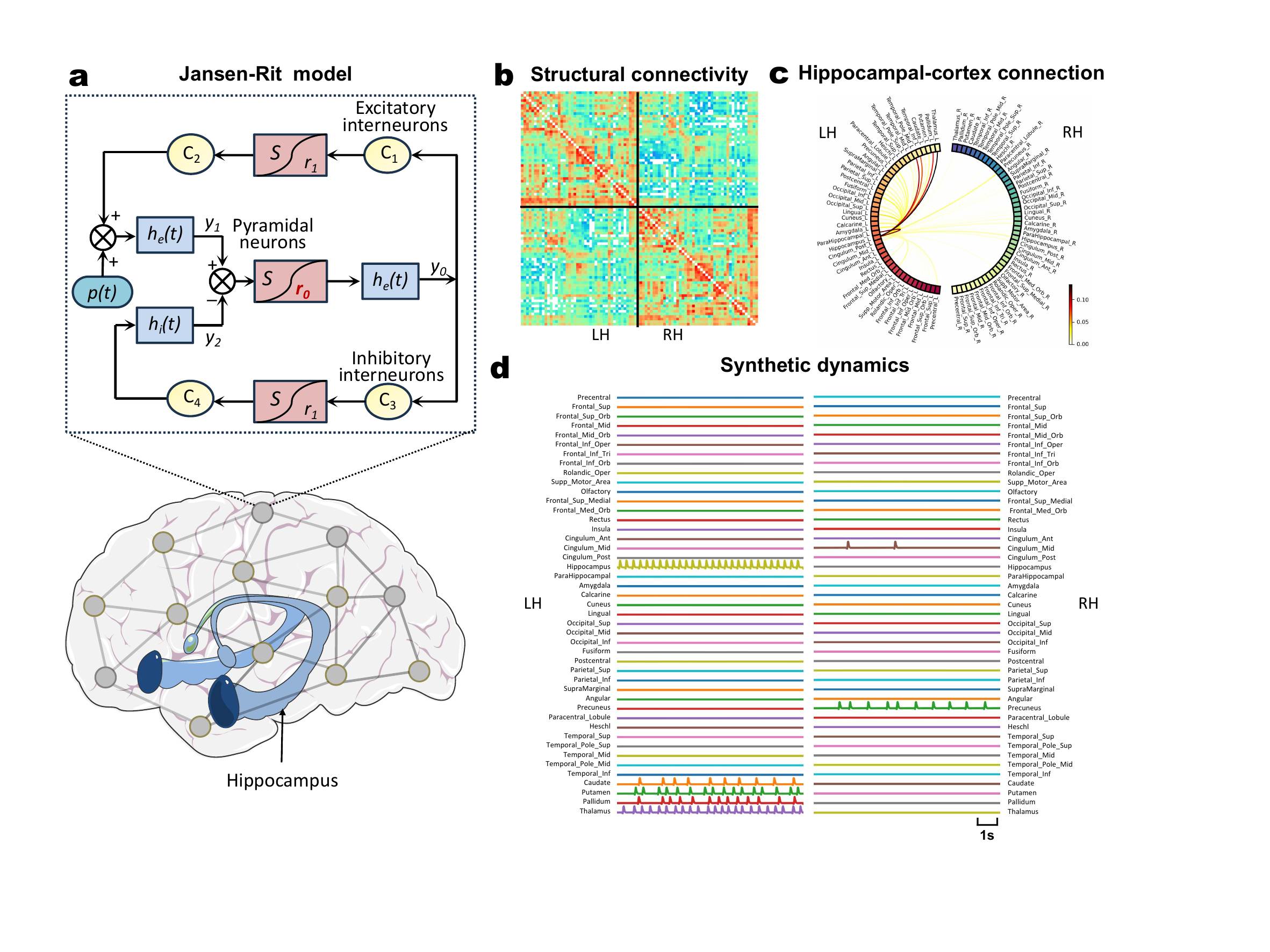}
\caption{\textbf{A simulation of the propagation process of seizure dynamics on the whole brain}. \textbf{a}, The neural dynamics of each node in the whole brain can be simplified as Jansen-Rit dynamics. The neural dynamics between interconnected regions are propagated through the white matter fiber connections. \textbf{b}, An example of the whole brain structural connectivity. \textbf{c}, The connection that links to the hippocampus. \textbf{d}, The simulated neural dynamics of the whole brain. The hippocampus is predefined as an SOZ with high-frequency abnormal discharge, while other regions with abnormal discharge are considered as propagation regions. The regions with normal discharge are non-seizure regions.} \label{fig:simulation}
\end{figure}

We initially applied the network-coupled Jansen-Rit model to simulate the dynamics of seizure propagation. The standard Jansen-Rit model delineates a cortical column that is composed of the complex functional interaction among excitatory, inhibitory interneurons, and pyramidal neurons (Fig.~\ref{fig:simulation}a)~\cite{Jansen_Rit_1995}. As shown in Fig.~\ref{fig:simulation}a, the functional dynamics of these three populations contain two modules: 

(1) The conversion of average pulse density into average postsynaptic membrane potential (PSP), which is described as the impulse response function for both excitatory and inhibitory interneurons:
\begin{equation}
    \begin{aligned}
        \left.h_E(t)=\left\{\begin{array}{ll}Aate^{-at},&t\geq0\\\\0,&t<0,\end{array}\right.\right.
    \end{aligned}
\end{equation}
and 
\begin{equation}
\begin{aligned}
h_I(t)=\left\{\begin{array}{ll}Bbte^{-bt},&t\ge0\\\\0,&t<0,\end{array}\right.
\end{aligned}
\end{equation}
where the constants $A$ and $B$ determine the maximum amplitude of the PSPs for excitatory (EPSPs) and inhibitory (IPSPs) responses respectively; $a$ and $b$ are the inverse time constants for the excitatory and inhibitory postsynaptic potentials respectively. 

(2) Then it translates the postsynaptic membrane potentials back into average pulse density, characterized by a sigmoid function:
\begin{equation}
\begin{aligned}
S(v,r)=\frac{\zeta_{max}}{1+e^{r(\theta-v)}},
\end{aligned}
\end{equation}
where $\zeta_{max}$ denotes the maximum firing rate, $r$ the slope of the function, and $\theta$ the half maximal response of the neuron population. Additionally, pyramidal neurons are stimulated externally by $p(t)$. 

Considering the information conversion with other interconnection cortical columns, we can model the neural dynamics of each cortical column as 
\begin{equation}
\begin{aligned}
\dot{x}_{0,i}(t) &=y_{0,i}(t) \\
\dot{y}_{0,i}(t) &=Aa\left[S(x_{1,i}(t)-x_{2,i}(t))\right]-2ay_{0,i}(t)-a^2x_{0,i}(t) \\
\dot{x}_{1,i}(t) &=y_{1,i}(t) + u_i(t))\\
\dot{y}_{1,i}(t) &=Aa\left[p(t)+C_2 S(C_1x_{0,i}(t)+g\sum_{j=1}^NM_{i,j}x_{3,j}\right]-2ay_{1,i}(t)-a^2x_{1,i}(t) \\
\dot{x}_{2,i}(t) &=y_{2,i}(t) \\
\dot{y}_{2,i}(t) &=Bb\left[C_4S(C_{3}x_{0,i}(t),r_{2})\right]-2by_{2,i}(t)-b^{2}x_{2,i}(t) \\
\dot{x}_{3,i}(t) &=y_{3,i}(t) \\
\dot{y}_{3,i}(t) &=A\bar{a}\left[S(x_{1,i}(t)-x_{2,i}(t))\right]-2\bar{a}y_{3,i}(t)-\bar{a}_{i}^{2}x_{3,i}(t) 
\end{aligned}
\end{equation}
where $x_{0}$, $x_{1}$, and $x_{2}$ represent the outputs of the PSP blocks for pyramidal, excitatory, and inhibitory neurons, respectively. $x_{3}$ is the output for long-range interactions of pyramidal neurons. The constants $C_{1}$, $C_{2}$, $C_{3}$, and $C_{4}$ scale the connectivity among the neural populations. 
$M_{i,j}$ is the structural connections between cortical column $i$ and $j$ (Fig.~\ref{fig:simulation}b). $g$ is the strength of global connections. $u_i(t)$ is the input of node $i$. The detailed information about the Jansen-Rit neural dynamics is summarized in study~\cite{Jansen_Rit_1995,coronel2021cholinergic}.


Considering that the hippocampus is a common SOZ in focal epilepsy, in our simulation we set the hippocampus as SOZ (Fig.~\ref{fig:simulation}a, bottom). We set the amplitude of PSPs for excitatory interneurons with $A=7.8$ for a seizure onset zone and $A=6.8$ for others. The seizure dynamics propagate to the whole brain through the structural connection (Fig.~\ref{fig:simulation}d). The high-frequency abnormal discharges are shown in the SOZ, and the low-frequency dynamics occur in the propagation regions with strong SC (Fig.~\ref{fig:simulation}d). 



\subsection{Real iEEG data from an epileptic patient}

To study whether the control strategy performs well on the real data, we conduct experiments on the real neural dynamical data from an epileptic patient. The real iEEG data is from an open dataset (Epilepsy-iEEG-Multicenter-Dataset,\url{https://openneuro.org/datasets/ds004100/versions/1.1.3}). 
Each subject recorded iEEG signals for multiple complete seizure events (from 120 seconds before seizure onset to 60 seconds after seizure termination) at a sampling rate exceeding 512 Hz.
To keep consistent with the JR simulation, we selected the patient (i.e., Subject HUP-146) with SOZ located in the hippocampus and with successful surgical outcomes, implying that the SOZ was correctly labeled by medical experts. In total 122-channel iEEG signals were recorded from this patient (Fig.~\ref{fig:framework} a). 

To be noted, we did not directly apply electrical stimulation to the patient. Instead, we first learn the parameters of the dynamical model in a data-driven fashion, by fitting the model with the real iEEG data. Then we conduct the virtual stimulation experiment on the identified dynamical model, identical to the experiments run on the simulation platform.



\subsection{Experiment protocol}

The selection of control nodes determines the control matrix $B$. For the control experiments on the simulation platform, we consider 4 types of control nodes to apply virtual stimulation, including (1) \textit{SOZ hub}, i.e., hippocampus, (2) \textit{SC hub}, the hub node with the strongest structural connection to SOZ, (3) \textit{FC hub}, the hub node with the strongest functional connectivity with SOZ, (4) \textit{Top weighted degree}, the node that connects to SOZ with the highest degree. 
Specifically, in the simulation platform, the hippocampus in the left hemisphere (LH) is the defined SOZ, the thalamus in LH is the hub node with the strongest structural connection and functional connection to the hippocampus, and the precuneus in RH has the highest degree of the whole brain and connects to SOZ. Towards these 4 types of control nodes in simulation, we solved the corresponding control problem formulated in Eq.~\eqref{Eq:MPC} separately. The optimized control strategy $u^*$ of Eq.~\eqref{Eq:MPC} was then applied to the network-coupled Jansen-Rit dynamics.


For the control experiment on the dynamical model from the real iEEG data, we consider 3 types of control nodes, including (1) \textit{SOZ}, in total 11 nodes which are predefined by medical experts, (2) \textit{FC hub}, the node with the highest FC with SOZ, (3) \textit{Top weighted degree}, the top-11 highest degrees of FC. The covariance-based functional connectivity is estimated on the first 2s of iEEG during the ictal period. In our experiment, it is not a direct electrical stimulation to the patient. The dimensionless variable $u(t)$ in the surrogate dynamics $x(t+1)=Ax(t)+Bu(t)$ has no physical meaning. We can obtain the optimized $u^*$ of Eq.~\eqref{Eq:MPC} iteratively. The optimized control strategy was applied to the surrogate system learned from iEEG data, rather than the real brain.

\begin{figure}[!th]
\centering
\includegraphics[width=0.95\textwidth]{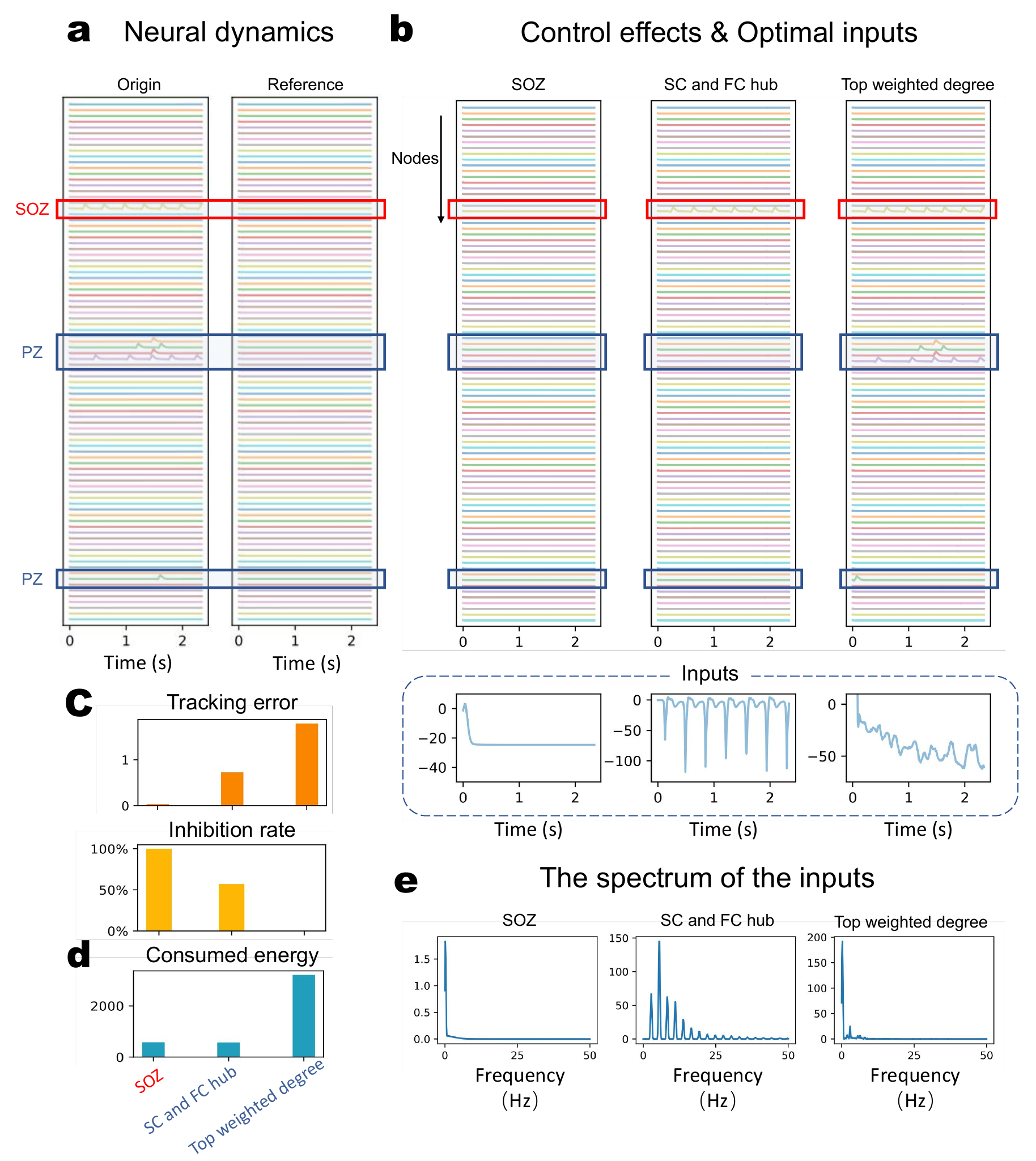}

\caption{\textbf{Control effects on the JR simulation platform}. We input the optimized electrical stimulation to the pre-selected control node and illustrate the control effects. \textbf{a}, The original seizure dynamics (left) and reference neural dynamics (right). \textbf{b}, Control effects (top) and the optimal inputs (bottom). From left to right, the control node is set to the hippocampus (SOZ), thalamus (SC and FC hub), and precuneus (Top weighted degree), respectively. The results show that stimulating the hippocampus or thalamus can successfully suppress seizure propagation. \textbf{c}, The mean absolute error of system dynamics and reference dynamics. \textbf{d}, The average energy consumed to control the system. \textbf{e}, The spectrum of optimal inputs on each pre-selected node.} 
\label{fig:JR_Simulation}
\end{figure}

\section{Results}

\subsection{Suppressing seizure on simulation platform}
In the simulation platform, we validate the sufficiency of seizure suppression with the hub control node on a network-coupled Jansen-Rit dynamics. We select one of the cortical columns (Hippocampus, with $A=7.8$ in Eq.~\eqref{fig:JR_Simulation}) as a SOZ to simulate the seizure propagation process. The neural dynamics are propagated through the white matter fiber tracts (the structural connections are projected in the AAL atlas). We also define the reference dynamics with $A=6.8$ in Eq.~\eqref{fig:JR_Simulation} of the whole brain (Fig.~\ref{fig:JR_Simulation} a).

We simulate the neural dynamics for 2.5 seconds with a 100 Hz sampling rate (Fig.~\ref{fig:JR_Simulation} a). We inserted inputs on the 4 types of pre-selected control nodes respectively (Fig.~\ref{fig:JR_Simulation} b). The results show that the optimal control strategy on SOZ, SC and FC hub have better control effects in suppressing seizure propagation than the node with the highest degree. When inserting optimal strategy on the hippocampus, both seizure-like activity on SOZ and PZ are suppressed. The optimal strategy on the thalamus suppressed the seizure-like activity on PZ and weakened the activity on SOZ.

To quantify the control effect, we further calculate the average absolute error between neural dynamics and reference after applying optimal stimulation. As well as the decrease in firing rate, \ie the ratio of the time points where the amplitude is suppressed to the total number of time points with high amplitude during epileptic discharge. It can be observed that when stimulating the hippocampus, the tracking error is the smallest and the inhibition rate of discharge reaches 100$\%$. 

The control effects on stimulating the thalamus showed a smaller tracking error with a global decrease rate of 57$\%$. The last type of control nodes have almost no control effect on suppressing epilepsy (see Fig.~\ref{fig:JR_Simulation} c). Considering the safety of electrical stimulation, to achieve the effect of reducing epileptic seizures while requiring lower power stimulation, we analyzed the control energy in Fig.~\ref{fig:JR_Simulation} d. The result shows that the energy of electrical stimulation applied to the hippocampus and the thalamus is much lower than the Precuneus. Interestingly, we found that the optimal input for the thalamus appears as a periodic pulse sequence. Further spectral analysis revealed that the primary frequency is concentrated around 5 Hz (see Fig.~\ref{fig:JR_Simulation} e). This low-frequency, low-energy stimulation pattern may be valuable for designing electrical stimulation in real-world applications. 


\subsection{Suppressing Real Seizure Dynamics}

\begin{figure}[!th]
\centering
\includegraphics[width=\textwidth]{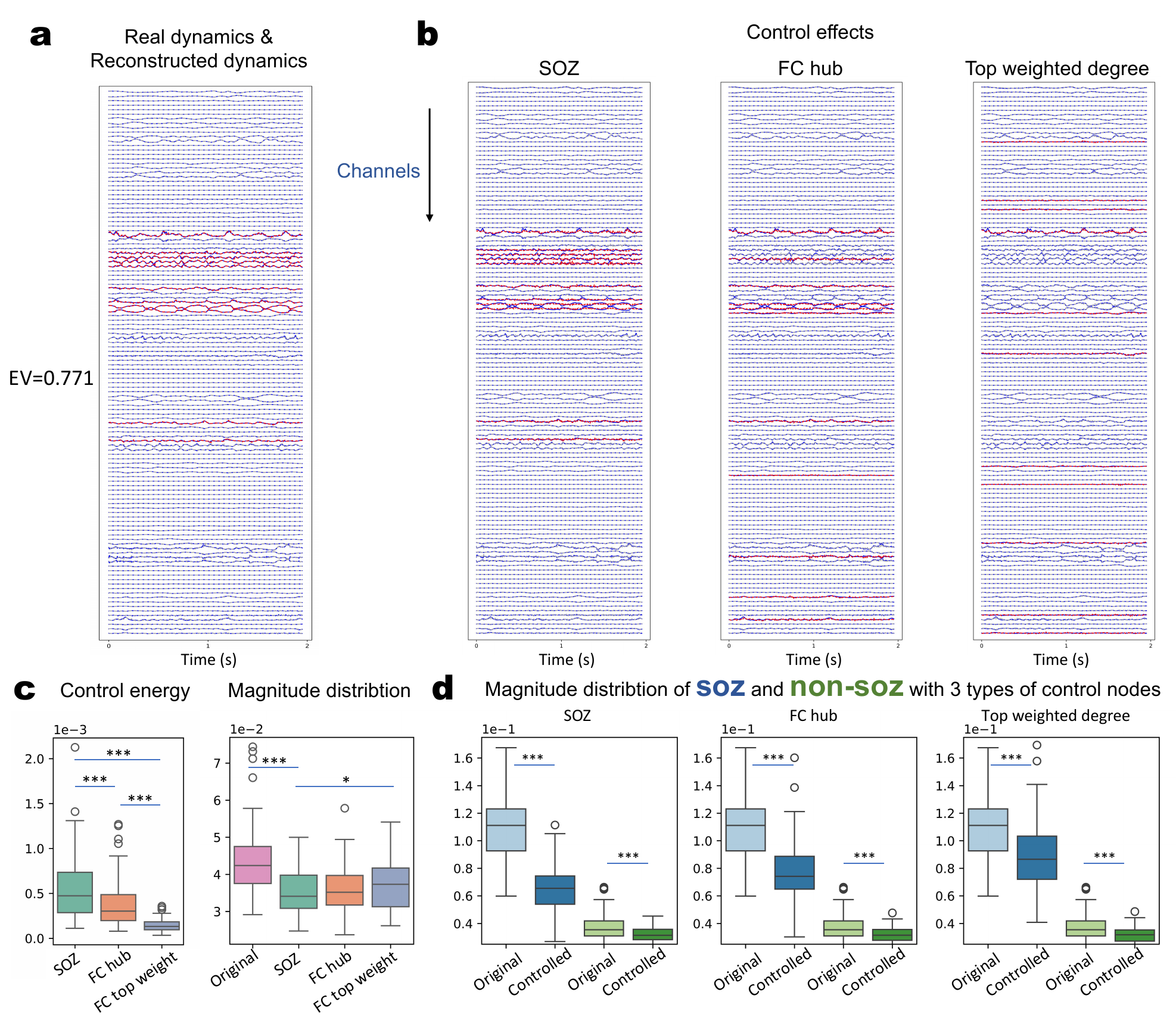}
\caption{\textbf{Control effects on real data}. We input the optimized virtual electrical stimulation to the pre-selected control node and illustrate the control effects on the identified model $x_{k+1}=Ax_{k}+Bu_k$. \textbf{a}, The original seizure dynamics (gray line) and the reconstructed neural dynamics (\textcolor{blue}{blue} dot line). \textbf{b}, Control effects with optimal inputs to three types of control nodes (left: SOZ, middle: top degree of FC, right: FC hub). \textbf{c}, Distribution of control energy (left) and magnitude (right). \textbf{d}, Magnitude distribution of soz (\textcolor{blue}{blue-paired}) and non-soz (\textcolor{teal}{green-paired}) with 3 types of control nodes. ($*$ for $p$-value$<0.05$ and $***$ for $p$-value$<1e^{-4}$)} 
\label{fig:real_data}
\end{figure}

We further validate the efficiency of seizure suppression on real data. We first validate the prediction performance of the DMD framework $x(t+1)=Ax(t)$ in predicting seizure dynamics. During the experiment, the hyperparameter of the length of the time window is $time\_length=512$, and the length of prediction is $n\_horizon=5$. In Fig.~\ref{fig:real_data} a, the result shows sufficiency prediction accuracy on real data with explained variance $EV=0.771$. Then we conduct the virtual stimulation experiment on the identified system with the consideration of control node selection and designing optimal control strategy. Control node selection determines the control matrix $B$, which is a binary diagonal matrix with the index controlled or not. 

Then we conduct experiments equipped with the system identification module on real data. The control effects of stimulating 3 types of control nodes (i.e., SOZ, FC hub, FC top weight) are shown in Fig.~\ref{fig:real_data} b respectively. The results show that both SOZ and FC hubs can suppress seizures.
To quantify the control effects, we calculated the energy of the input sequences and the amplitude of the neural signals influenced by the inputs, respectively. The quantitative results are shown in Fig.~\ref{fig:real_data}c. We show that the control to SOZ requires the highest input energy, while it has the maximal neural magnitude reduction, indicating effective suppression of seizures. On the contrary, the inputs to FC top weight are inefficient in suppressing neural activity. These two control nodes showed significant statistical differences in both control energy and amplitude distribution ($p$-value$<1e^{-4}$ of two-sided t-test). There are no significant differences between controlling SOZ and controlling the FC hub in amplitude reduction of neural activity. 

Furthermore, we compared the control effects on SOZ and non-SOZ given the optimal inputs with 3 types of control node selection, as shown in Fig.~\ref{fig:real_data} d. The results show that both the SOZ and non-SOZ decrease the amplitude of neural activity compared with the original neural activity with significant statistical differences ($p$-value$<10^{-4}$ of two-sided t-test). However, the control effects with optimal inputs on control nodes that are related to SOZ are sufficient to suppress seizure dynamics.


\section{Discussion and conclusion}
\subsubsection*{From methodology perspective}
Our study presented a general platform based on system identification and brain network control theory for suppressing seizure dynamics. We applied the dynamic mode decomposition to iteratively update the neural dynamics online, which guarantees high predictive performance in reconstructing neural dynamics. The platform was validated with both synthetic data and real iEEG data from the Epilepsy-iEEG-Multicenter-Dataset. With the high predictive performance, we further conduct model predictive control with the identified surrogate model in inferring the optimal control strategy under the consideration of different types of control nodes. However, the sparsity of sEEG affects the system identification of the whole brain. Lou et al proposed the framework of fusion multi-modal neural dynamics to reconstruct the whole brain neural dynamics using sEEG-EEG data, which can improve the performance of system identification~\cite{lou2024data}.

\subsubsection*{From clinical perspective}

Validating the effectiveness of different stimulation strategies on neural treatment is the core question in clinical neuroscience. Our work delves into the optimal electrical stimulation to the hub of the epileptic brain network, focusing on two key aspects: control node selection and inferring optimal inputs. Selecting the control node enables the validation of the feasibility of conducting the electrical stimulation to reach target states. Moreover, the optimal inputs are key to realizing the objective with less consumed energy. 

In summary, we proposed a platform from the control theory perspective for validating the efficiency of different control strategies (control node selection) in suppressing seizures. The hub of the epileptic brain network serves as the alternative control node for realizing seizure suppression.

\section*{Acknowledgement}
The authors gratefully acknowledge Prof. Xiang Liao, Dr. Liang Chen and Dr. Chen Yao for their insightful discussions, comments and suggestions. This work was funded in part by the National Key R\&D Program of China (2021YFF1200804), Shenzhen Science and Technology Innovation Committee (2022410129, KCXFZ20201221173400001, KJZD20230923115221044).

\bibliography{mybibliography}
\bibliographystyle{unsrt}
\end{sloppypar}
\end{document}